\documentclass[10pt,conference]{IEEEtran}

\usepackage[utf8]{inputenc}
\usepackage{times}
\usepackage{graphicx}
\usepackage{amsmath}
\usepackage{amsfonts}
\usepackage{amssymb}
\usepackage{mathtools}
\usepackage{hyperref}
\usepackage{booktabs}
\usepackage{algorithm}
\usepackage{algorithmic}
\usepackage{xcolor}
\usepackage{url}
\usepackage{float}

\graphicspath{{figures/}}

\definecolor{linkblue}{RGB}{0,75,150}
\hypersetup{
   colorlinks=true,
   linkcolor=linkblue,
   urlcolor=linkblue,
   citecolor=linkblue,
 }

\begin{document}

\title{Diagonal Scaling: A Multi-Dimensional Resource Model \\
and Optimization Framework for Distributed Databases}

\author{
    Shahir Abdullah \\ 
    \texttt{shahir2561@gmail.com}
    \and
    Syed Rohit Zaman \\
    \texttt{0423052004@grad.cse.buet.ac.bd}
}

\maketitle

\begin{abstract}
Modern cloud databases usually expose elasticity through two independent mechanisms: scale-out by adding nodes and scale-up by increasing per-node resources. This binary view is limiting because latency, throughput, cost, and coordination overhead arise from the joint interaction between cluster size and per-node capacity. This paper introduces the \emph{Scaling Plane}, a two-dimensional resource model in which a configuration is represented as $(H,V)$, where $H$ is the number of nodes and $V$ is a vertical resource tier containing CPU, memory, network, and storage capacity.

We formalize analytical surfaces for latency, throughput, coordination cost, cluster cost, and a composite objective function. Based on this model, we propose \textsc{DiagonalScale}, a local-search autoscaling policy that evaluates horizontal, vertical, and diagonal neighbors in the Scaling Plane. Unlike earlier axis-aligned policies, \textsc{DiagonalScale} explicitly filters infeasible configurations using SLA constraints and selects the lowest-cost feasible neighbor after applying a rebalance penalty.

We evaluate the current prototype using a Phase-1 analytical simulator over 16 cluster configurations, four vertical tiers, and a 50-step dynamic workload trace. The latest simulation shows that \textsc{DiagonalScale} reduces SLA violations to 3 out of 50 steps, compared with 32 for horizontal-only scaling and 21 for vertical-only scaling. It also achieves the lowest average latency (4.05 synthetic latency units) and the lowest average objective value (65.53), at a comparable but slightly higher average cost. These results provide early evidence that diagonal scaling can achieve better latency--SLA trade-offs than traditional horizontal-only or vertical-only policies. The current evaluation is intentionally analytical; future work will calibrate the surfaces with real distributed database benchmarks.
\end{abstract}

\section{Introduction}

Distributed databases power mission-critical services such as financial systems, e-commerce platforms, logistics backends, learning platforms, and globally replicated applications. Their practical value depends not only on correctness and availability, but also on elastic behavior: the system must absorb load changes while maintaining acceptable latency and cost. Cloud-native databases usually expose two scaling actions. Horizontal scaling changes the number of nodes, while vertical scaling changes the resources available to each node. These actions are often configured separately, for example through horizontal autoscaling policies, instance-size changes, or manual operator decisions.

This separation is convenient, but it hides a deeper resource-management problem. Horizontal scaling improves parallelism and available aggregate throughput, but it can increase coordination latency, replication overhead, rebalancing cost, and metadata movement. Vertical scaling improves per-node capacity and can reduce local bottlenecks, but it does not increase the number of independent execution or storage partitions. As a result, neither dimension alone captures the full behavior of distributed databases such as Spanner~\cite{corbett2013spanner}, Bigtable~\cite{chang2008bigtable}, Dynamo~\cite{decandia2007dynamo}, Cassandra~\cite{lakshman2010cassandra}, CockroachDB~\cite{cockroachdb_docs}, and YugabyteDB~\cite{yugabyte_docs}.

This work proposes that database elasticity should be viewed as movement through a two-dimensional resource space rather than as a binary choice. We call this space the \emph{Scaling Plane}. Each point in the plane represents a deployable configuration $(H,V)$, where $H$ is the number of nodes and $V$ is a vertical resource tier. Over this plane, one can define latency, throughput, monetary cost, coordination cost, and composite objective surfaces. These surfaces make it possible to reason geometrically about when a system should scale out, scale up, scale down, or move diagonally.

\subsection{Motivation}

Traditional autoscalers often rely on simple thresholds: scale out when CPU usage crosses a boundary, scale up when memory pressure persists, or add replicas when request rate increases. Such policies are easy to implement, but they do not directly reason about the interaction between dimensions. A horizontal-only policy may keep adding nodes even when stronger per-node resources would avoid repeated rebalancing. A vertical-only policy may use expensive instances while failing to exploit available workload parallelism. A mixed policy without a formal objective may oscillate between these behaviors.

The central observation behind this paper is that many scaling decisions are not purely horizontal or purely vertical. A cluster may require both a moderate increase in node count and a moderate increase in per-node capacity. Conversely, a system may need to reduce one dimension while holding the other fixed. The term \emph{diagonal scaling} captures such coordinated movement through the Scaling Plane.

\subsection{Research Questions}

This paper studies the following questions:

\begin{enumerate}
    \item How can distributed database configurations be represented in a joint horizontal--vertical resource space?
    \item How can latency, throughput, cost, and coordination overhead be modeled over that space?
    \item Can a local-search policy over horizontal, vertical, and diagonal neighbors outperform axis-aligned autoscaling baselines?
    \item What does an analytical Phase-1 simulation reveal about the strengths and weaknesses of diagonal scaling?
\end{enumerate}

\subsection{Contributions}

The paper makes the following contributions:

\begin{enumerate}
    \item \textbf{Scaling Plane abstraction.} We introduce a two-dimensional model for reasoning about distributed database elasticity.
    \item \textbf{Analytical surfaces.} We define latency, throughput, coordination-cost, cluster-cost, and objective surfaces over $(H,V)$.
    \item \textbf{SLA-aware DiagonalScale policy.} We propose a local-search autoscaling policy that evaluates horizontal, vertical, and diagonal neighbors while filtering infeasible configurations.
    \item \textbf{Phase-1 simulator.} We implement an analytical simulator that evaluates all configurations, produces heatmaps and surfaces, and compares autoscaling policies over a dynamic workload trace.
    \item \textbf{Updated simulation findings.} The latest simulation shows that \textsc{DiagonalScale} achieves the lowest average latency and objective value, and reduces SLA violations from 32 and 21 for the baselines to 3.
\end{enumerate}

\section{Background and Related Work}

\subsection{Horizontal Scaling}

Horizontal scaling increases the number of nodes in a cluster. This is the dominant approach in many distributed databases. Spanner uses globally replicated data and consensus to support external consistency~\cite{corbett2013spanner}; Bigtable partitions data into tablets and distributes them across tablet servers~\cite{chang2008bigtable}; Dynamo focuses on highly available key-value storage using consistent hashing and replication~\cite{decandia2007dynamo}; Cassandra similarly uses a decentralized architecture and tunable consistency~\cite{lakshman2010cassandra}.

The benefit of horizontal scaling is increased aggregate capacity and fault tolerance. The cost is that more nodes can introduce more network communication, metadata management, replica coordination, shard movement, and rebalance operations. Strongly consistent systems may pay additional costs for quorum protocols, while eventually consistent systems may incur anti-entropy or repair costs. These effects motivate including a coordination term in the objective function.

\subsection{Vertical Scaling}

Vertical scaling increases the CPU, memory, network bandwidth, or storage IOPS assigned to each node. This can reduce queueing, improve cache locality, accelerate compaction, and increase per-node throughput. However, vertical scaling has limits. Cloud instance families are discrete, costs often rise sharply with instance size, and a single stronger node does not provide the same partition-level parallelism as multiple nodes. Kubernetes VPA~\cite{k8s_vpa} and cloud database services such as Aurora Serverless v2~\cite{aurora_serverlessv2} provide mechanisms that are related to vertical elasticity, but the interaction between horizontal and vertical dimensions is not usually formalized as a joint optimization problem.

\subsection{Autoscaling and Resource Management}

Kubernetes HPA~\cite{k8s_hpa} changes the number of replicas, while VPA~\cite{k8s_vpa} adjusts requested resources. Cluster Autoscaler~\cite{k8s_cluster_autoscaler} changes the number of worker nodes. Large-scale resource management systems such as Borg~\cite{verma2015large} show that cluster management is a central systems problem. However, these systems typically manage compute resources broadly, whereas our work focuses on the specific interaction between cluster size, database coordination, and per-node resource capacity.

\subsection{Gap Addressed by This Work}

Prior work has studied scalable storage systems, cloud resource management, and serverless computing~\cite{hellerstein2015serverless,armbrust2010view}. The gap addressed here is not the absence of horizontal or vertical scaling mechanisms. Instead, the gap is the absence of a simple joint model that treats database scaling as movement through a structured resource plane. The Scaling Plane is intended as an abstraction that can support analytical simulation, empirical calibration, and future learned controllers.

\section{Scaling Plane Model}
\label{sec:scaling-plane}

\subsection{Configuration Space}

A database configuration is represented as:
\[
(H,V),
\]
where $H \in \mathbb{N}$ is the number of nodes and $V$ is a vertical resource tier. In the current simulator, $V$ is selected from:
\[
V \in \{\text{small},\text{medium},\text{large},\text{xlarge}\}.
\]
Each tier contains CPU, RAM, network bandwidth, storage IOPS, and hourly cost. The simulation uses:
\[
H \in \{1,2,4,8\}.
\]
Thus the Phase-1 configuration space contains $4 \times 4 = 16$ points.

\subsection{Cost Surface}

\begin{figure}[t]
\centering
\includegraphics[width=\linewidth]{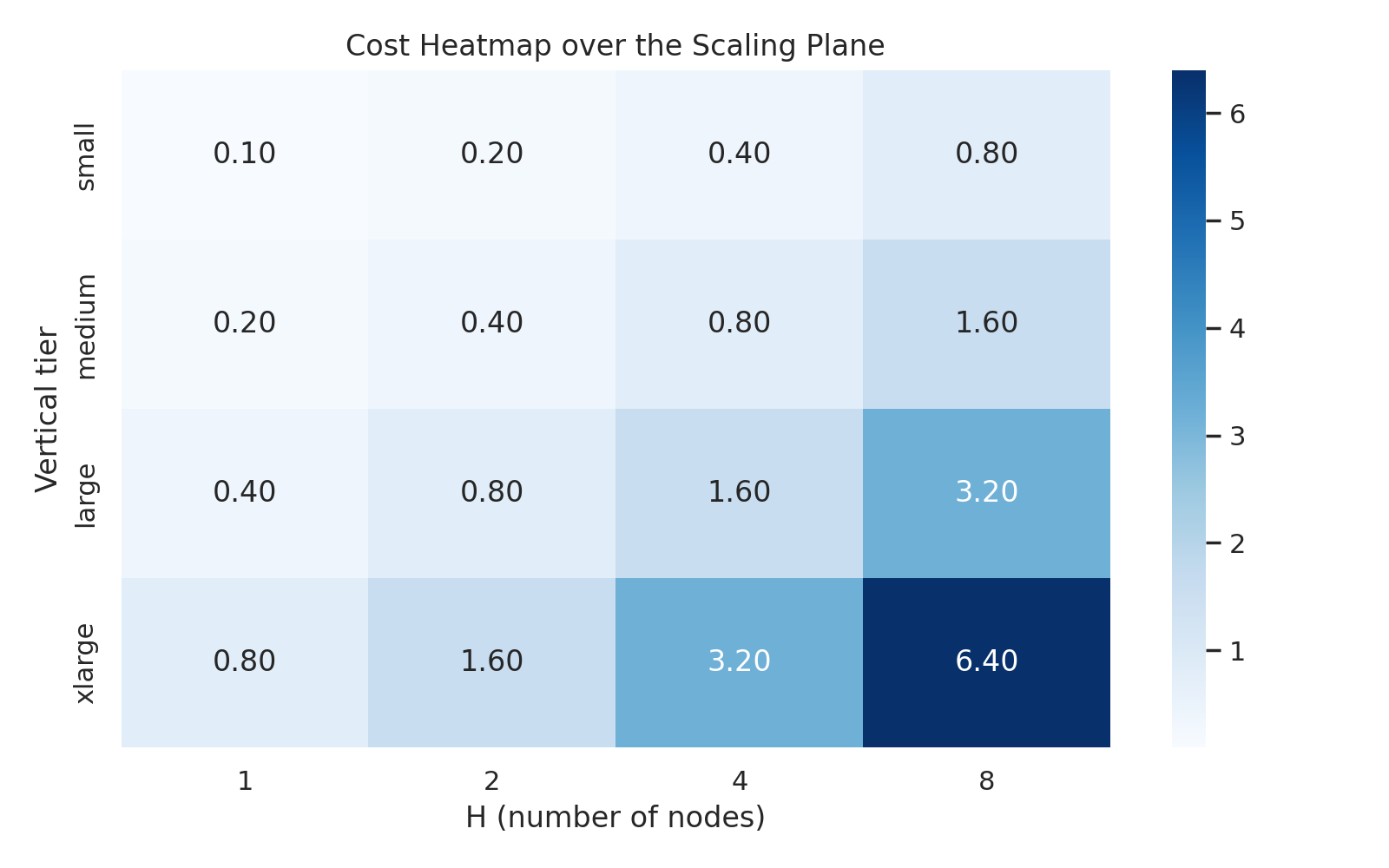}
\caption{Cost heatmap over the Scaling Plane. The cost surface is monotonic in both dimensions: increasing either the number of nodes $H$ or the vertical resource tier $V$ increases total cluster cost. This figure validates the cost component used by the simulator.}
\label{fig:cost-heatmap}
\end{figure}

The cluster cost is modeled as:
\[
C(H,V)=H \cdot C_{\text{node}}(V).
\]
The cost heatmap in Figure~\ref{fig:cost-heatmap} shows the expected monotonic behavior: cost increases with both node count and vertical tier.

\subsection{Latency Surface}

\begin{figure}[t]
\centering
\includegraphics[width=\linewidth]{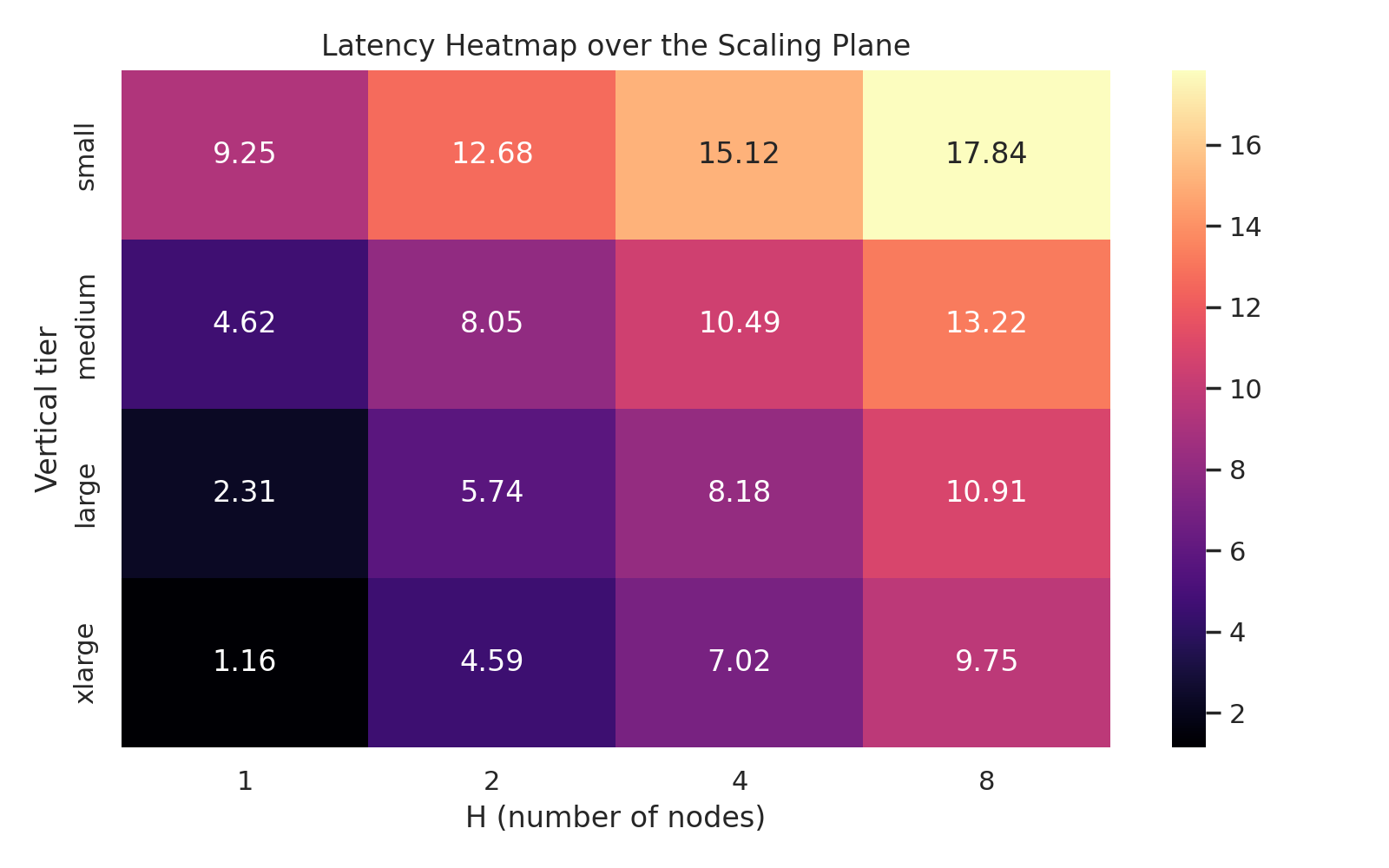}
\caption{Latency heatmap over the Scaling Plane. Latency decreases as vertical resources improve from small to xlarge, but increases as the number of nodes grows because of the modeled coordination term. This produces the core trade-off motivating diagonal scaling.}
\label{fig:latency-heatmap}
\end{figure}

The simulator decomposes latency into node-intrinsic latency and coordination latency:
\[
L(H,V)=L_{\text{node}}(V)+L_{\text{coord}}(H).
\]
The node term decreases with larger resource tiers:
\[
L_{\text{node}}(V)=\frac{a}{cpu}+\frac{b}{ram}+\frac{c}{bandwidth}+\frac{d}{iops/1000}.
\]
The coordination term increases with the number of nodes:
\[
L_{\text{coord}}(H)=\eta \log H + \mu H^\theta.
\]
This creates a surface where vertical scaling reduces node latency, while horizontal scaling increases coordination latency. Figure~\ref{fig:latency-heatmap} shows this effect clearly: larger vertical tiers reduce latency for a fixed $H$, while larger $H$ increases latency for a fixed tier.

\begin{figure}[t]
\centering
\includegraphics[width=\linewidth]{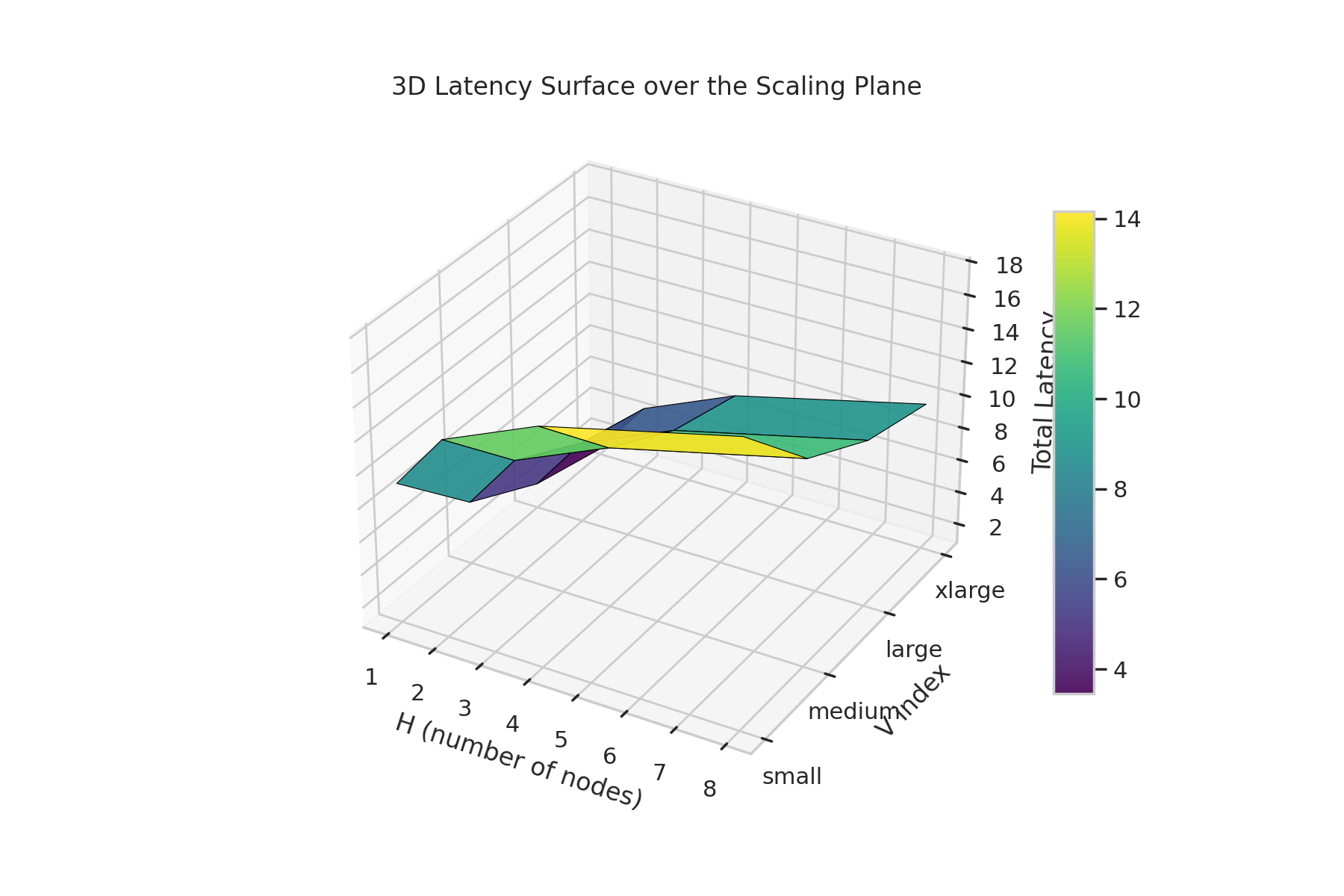}
\caption{3D latency surface over the Scaling Plane. The surface visualizes the opposing effects of vertical resources and horizontal coordination overhead. The surface slopes downward with larger vertical tiers and upward with larger node counts.}
\label{fig:latency-surface-3d}
\end{figure}

\subsection{Throughput Surface}

Throughput is modeled as:
\[
T(H,V)=H \cdot T_{\text{node}}(V) \cdot \phi(H),
\]
where:
\[
T_{\text{node}}(V)=\kappa \cdot \min(cpu,ram,bandwidth,iops/1000)
\]
and:
\[
\phi(H)=\frac{1}{1+\omega \log H}.
\]
The factor $\phi(H)$ captures diminishing returns from horizontal scaling. Larger clusters increase aggregate capacity, but less than linearly because coordination and communication overhead grow with cluster size.

\subsection{Coordination Cost}

Write-heavy workloads increase the cost of coordination. The simulator therefore models coordination cost as:
\[
K(H,V)=\rho \cdot L_{\text{coord}}(H) \cdot \frac{\lambda_w}{T(H,V)},
\]
where $\lambda_w$ is the workload write arrival rate. This term penalizes large clusters under write pressure, while allowing high-throughput configurations to absorb coordination overhead more effectively.

\subsection{Objective Function}

\begin{figure}[t]
\centering
\includegraphics[width=\linewidth]{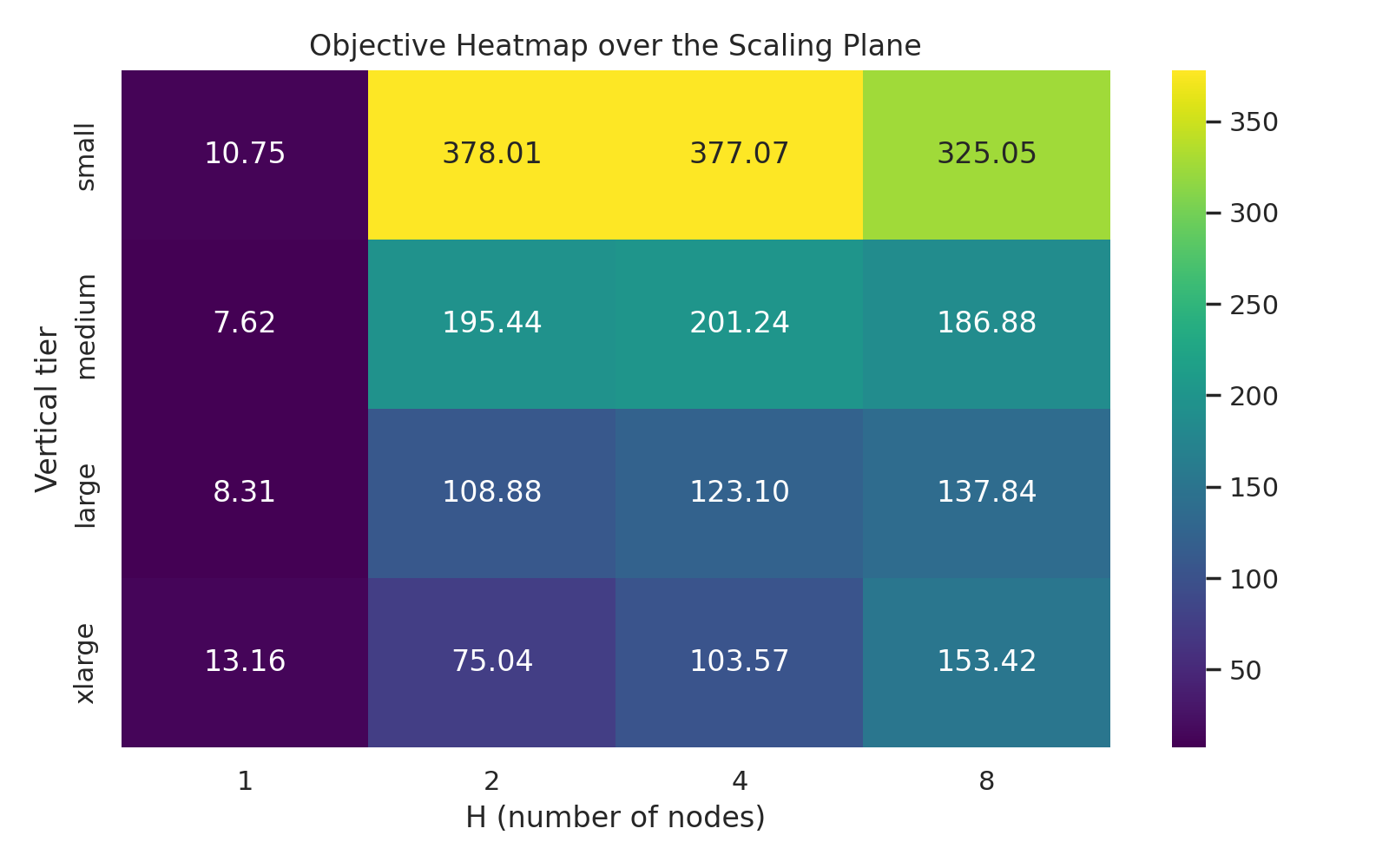}
\caption{Objective heatmap over the Scaling Plane for the default mixed workload. The unconstrained objective favors low-cost and low-coordination regions, but the dynamic policy simulation additionally enforces throughput and latency feasibility. This distinction is important: an objective-only minimum may be cheap but insufficient under high demand.}
\label{fig:objective-heatmap}
\end{figure}

The unconstrained objective used for surface exploration is:
\[
F(H,V)=\alpha L(H,V)+\beta C(H,V)+\gamma K(H,V)-\delta T(H,V).
\]
Latency, cost, and coordination cost are penalties; throughput is a reward. Figure~\ref{fig:objective-heatmap} shows the resulting objective landscape for the default mixed workload. The heatmap is useful for understanding how the model scores static configurations, but policy decisions also enforce SLA feasibility.

\section{DiagonalScale Algorithm}
\label{sec:algorithm}

\subsection{Policy Intuition}

\textsc{DiagonalScale} is a local-search policy over the Scaling Plane. At each time step, it observes the current configuration $(H,V)$ and current workload, generates neighboring configurations, removes those that violate SLA constraints, applies a rebalance penalty, and chooses the best remaining configuration. The key difference from horizontal-only and vertical-only baselines is that diagonal neighbors are first-class candidates.

\subsection{Neighbor Generation}

For a current configuration $(H,V)$, the policy evaluates:
\[
(H_{prev},V),\quad (H,V),\quad (H_{next},V),
\]
\[
(H,V_{prev}),\quad (H,V_{next}),
\]
and diagonal combinations such as:
\[
(H_{next},V_{next}), \quad (H_{prev},V_{prev}).
\]
In implementation, the neighbor set is generated using the previous and next valid values in the discrete $H$ and $V$ lists. This avoids invalid configurations and keeps the search space small.

\subsection{SLA Feasibility}

A candidate configuration is rejected if it is under pressure. The simulator checks two conditions:
\[
L(H,V) > L_{max},
\]
or:
\[
T(H,V) < \lambda_{req} \cdot b_{sla},
\]
where $\lambda_{req}$ is the workload-derived throughput requirement and $b_{sla}$ is a throughput buffer. This feasibility check was critical in the latest update. Without it, \textsc{DiagonalScale} could choose very cheap configurations that minimize the objective but fail throughput requirements.

\subsection{Rebalance Penalty}

To discourage disruptive moves, the adjusted score is:
\[
F'(H',V')=F(H',V')+R(H,V,H',V'),
\]
where:
\[
R=2|H'-H|+|V'-V|.
\]
Changing $H$ is penalized more heavily than changing $V$ because changing the node count can imply data movement, shard reassignment, and additional rebalancing.

\subsection{Algorithm}

\begin{algorithm}[t]
\caption{SLA-aware \textsc{DiagonalScale}}
\begin{algorithmic}[1]
\STATE Input: current configuration $(H,V)$ and workload $W$
\STATE Generate neighbors $\mathcal{N}(H,V)$
\STATE $best \leftarrow \emptyset$, $bestScore \leftarrow \infty$
\FOR{each $(H',V') \in \mathcal{N}(H,V)$}
    \STATE Compute $L(H',V')$, $T(H',V')$, $C(H',V')$, $K(H',V')$
    \IF{$L(H',V') > L_{max}$ or $T(H',V') < T_{min}$}
        \STATE continue
    \ENDIF
    \STATE Compute $F(H',V')$
    \STATE Compute rebalance penalty $R(H,V,H',V')$
    \STATE $score \leftarrow F(H',V') + R(H,V,H',V')$
    \IF{$score < bestScore$}
        \STATE $best \leftarrow (H',V')$
        \STATE $bestScore \leftarrow score$
    \ENDIF
\ENDFOR
\IF{$best = \emptyset$}
    \STATE return one-step diagonal scale-up fallback
\ELSE
    \STATE return $best$
\ENDIF
\end{algorithmic}
\end{algorithm}

\subsection{Complexity}

The policy evaluates at most nine candidates: the current configuration, horizontal neighbors, vertical neighbors, and diagonal neighbors. Since each candidate evaluation uses closed-form functions, the decision complexity is:
\[
O(|\mathcal{N}|)=O(1).
\]
This makes the policy suitable for a real-time control loop once the surface model has been calibrated.

\section{Simulation Methodology}
\label{sec:simulation}

\subsection{Simulator Scope}

The current results come from a Phase-1 analytical simulator. The simulator does not yet execute workloads against a live distributed database. Instead, it evaluates mathematical surfaces over a discrete configuration space and simulates policy decisions over a changing workload trace. This is the appropriate first step for testing whether the Scaling Plane abstraction and DiagonalScale policy behave coherently before investing in real-cluster benchmarking.

\subsection{Configuration Space}

The simulator uses:
\[
H \in \{1,2,4,8\}
\]
and four vertical tiers: small, medium, large, and xlarge. The cost heatmap in Figure~\ref{fig:cost-heatmap} shows the cost associated with these configurations.

\subsection{Workload Timeline}

The dynamic experiment uses a 50-step workload timeline:

\begin{itemize}
    \item steps 0--9: low intensity, $60$,
    \item steps 10--19: medium intensity, $100$,
    \item steps 20--29: high intensity, $160$,
    \item steps 30--39: medium intensity, $100$,
    \item steps 40--49: low intensity, $60$.
\end{itemize}

The workload is a mixed workload with read ratio $0.7$ and write ratio $0.3$. Required throughput is derived from workload intensity using a fixed scaling factor. The average required throughput across the trace is 9600 synthetic operations per unit interval.

\subsection{Policies Compared}

We compare three policies:

\begin{itemize}
    \item \textbf{Horizontal-only.} This policy changes only $H$ while keeping $V$ fixed.
    \item \textbf{Vertical-only.} This policy changes only $V$ while keeping $H$ fixed.
    \item \textbf{DiagonalScale.} This policy evaluates horizontal, vertical, and diagonal neighbors while enforcing SLA feasibility.
\end{itemize}

\subsection{Metrics}

The simulator reports average latency, maximum latency, average throughput, average required throughput, average cost, total cost, average objective value, and SLA violations. SLA violations are decomposed into latency violations and throughput violations.

\section{Results}
\label{sec:results}

\subsection{Summary of Latest Simulation}

Table~\ref{tab:policy-summary} reports the latest simulation output. \textsc{DiagonalScale} achieves the best average latency and objective score while nearly eliminating SLA violations. Horizontal-only scaling has high latency because increasing $H$ increases coordination cost. Vertical-only scaling has good latency, but it still produces many throughput violations because it is limited to one axis of adjustment.

\begin{table}[t]
\centering
\caption{Policy summary from the latest Phase-1 simulation. Lower latency, lower objective, and fewer SLA violations are better. Throughput values are synthetic simulator units.}
\label{tab:policy-summary}
\resizebox{\linewidth}{!}{%
\begin{tabular}{lrrrrrr}
\toprule
Policy & Avg. Lat. & Avg. Thr. & Avg. Cost & Total Cost & Avg. Obj. & SLA Viol. \\
\midrule
DiagonalScale & 4.05 & 13506.13 & 1.624 & 81.2 & 65.53 & 3 \\
Horizontal-only & 13.06 & 10293.20 & 1.560 & 78.0 & 180.94 & 32 \\
Vertical-only & 4.89 & 12068.66 & 1.416 & 70.8 & 77.70 & 21 \\
\bottomrule
\end{tabular}}
\end{table}

The key result is not that \textsc{DiagonalScale} is always cheapest. In fact, it pays slightly higher average cost than the baselines. The important point is that it spends that cost where it is useful: to satisfy the throughput SLA and reduce latency. Compared with horizontal-only scaling, \textsc{DiagonalScale} reduces average latency from 13.06 to 4.05 and reduces SLA violations from 32 to 3. Compared with vertical-only scaling, it reduces SLA violations from 21 to 3 while achieving lower average latency.

\subsection{Policy Trajectories}

\begin{figure}[t]
\centering
\includegraphics[width=\linewidth]{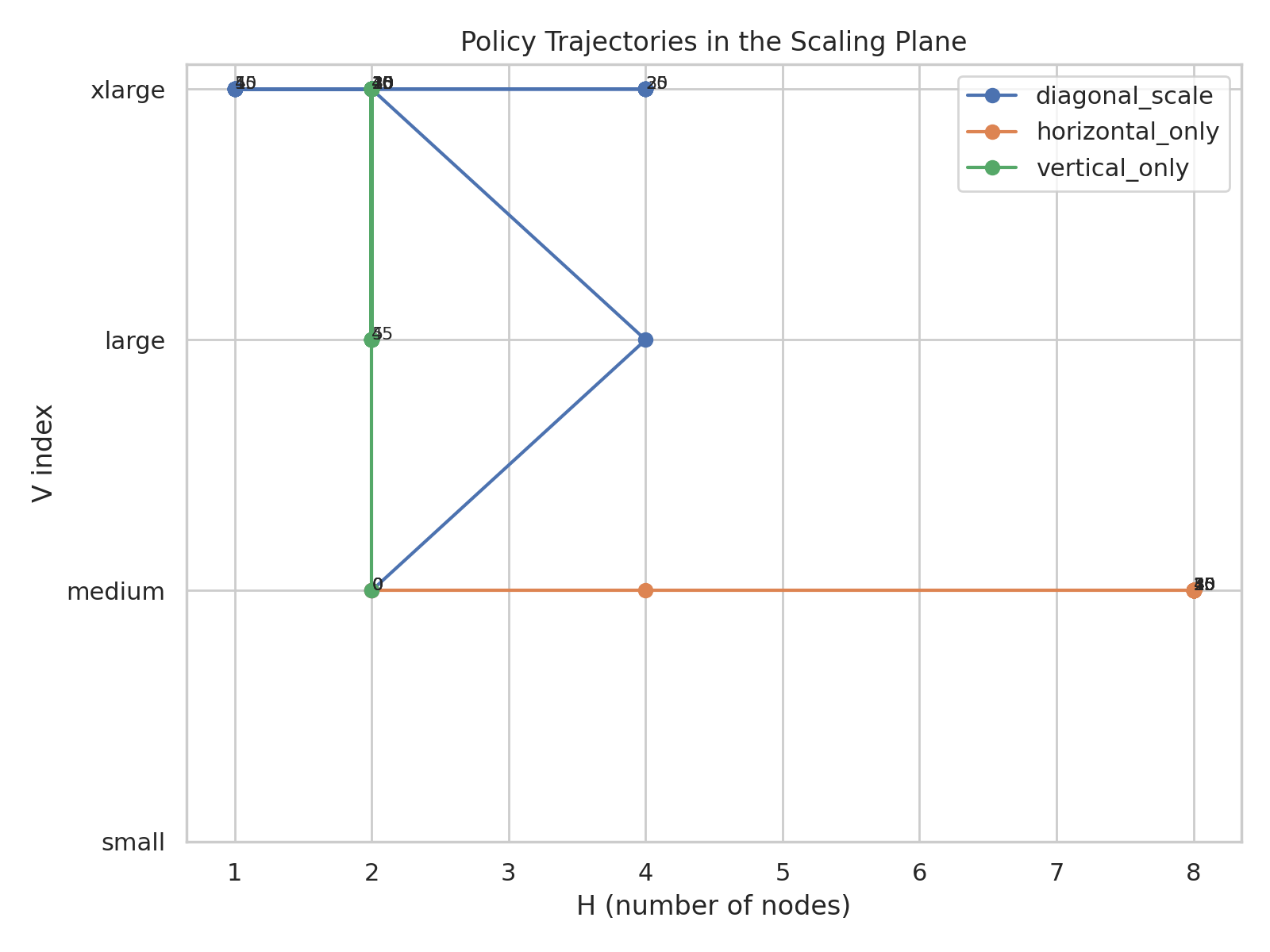}
\caption{Policy trajectories in the Scaling Plane. Horizontal-only moves along the $H$ axis, vertical-only moves along the $V$ axis, and \textsc{DiagonalScale} combines both types of movement. The trajectory demonstrates that the policy is not forced to move diagonally at every step; instead, it selects the best feasible neighbor at each decision point.}
\label{fig:policy-trajectories}
\end{figure}

Figure~\ref{fig:policy-trajectories} shows how the three policies move through the Scaling Plane. Horizontal-only scaling remains on the medium tier and moves primarily along the $H$ axis. Vertical-only scaling remains at a fixed node count and moves along the tier axis. \textsc{DiagonalScale}, in contrast, moves across both dimensions. During high load, it reaches stronger configurations; during lower load, it returns to cheaper configurations.

\subsection{Latency Over Time}

\begin{figure}[t]
\centering
\includegraphics[width=\linewidth]{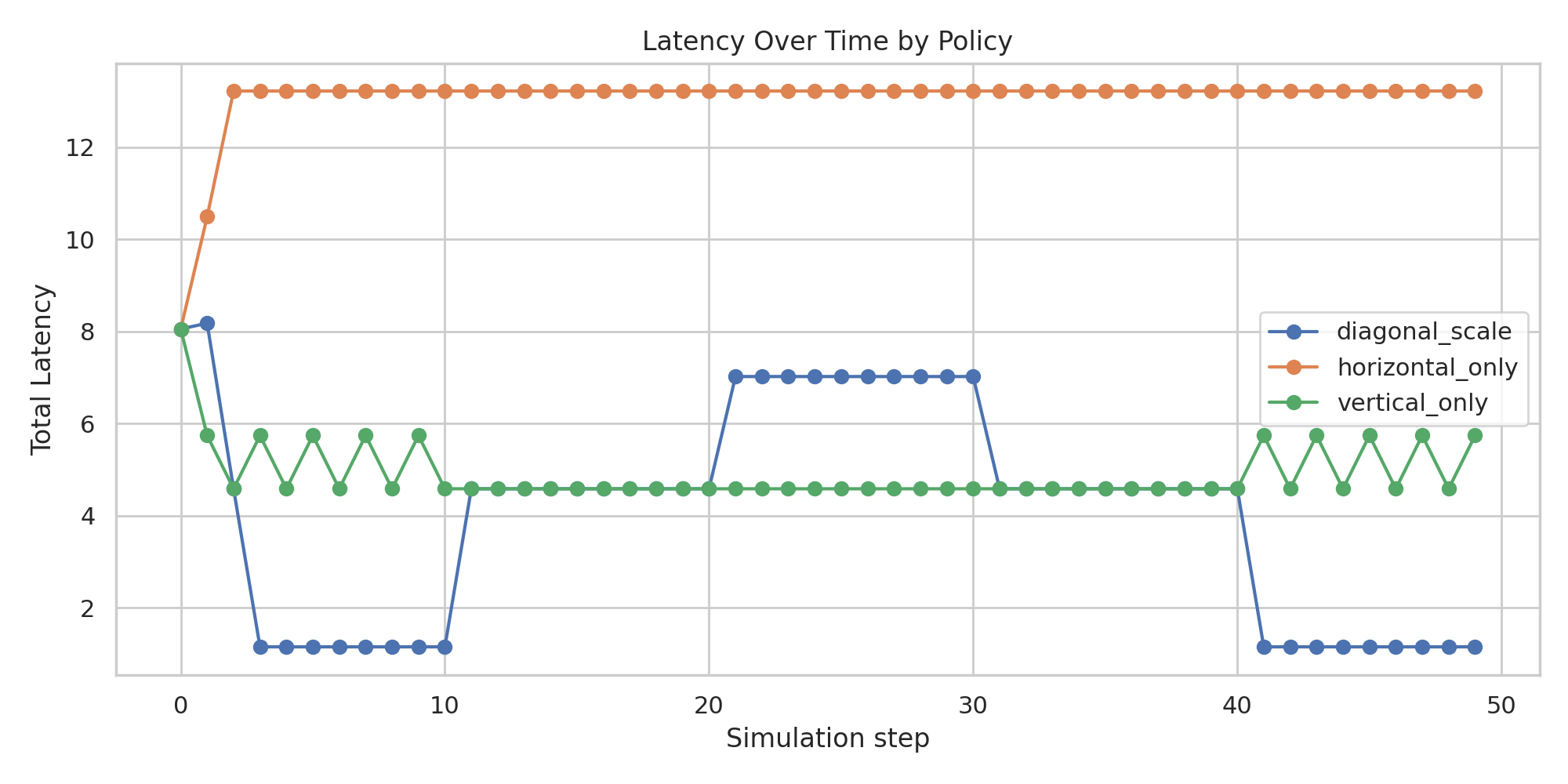}
\caption{Latency over time by policy. The horizontal-only policy remains at high latency because it adds nodes without improving per-node resources. Vertical-only scaling reduces latency but still lacks enough flexibility under some workload phases. \textsc{DiagonalScale} achieves the lowest average latency by combining both dimensions.}
\label{fig:policy-latency-time}
\end{figure}

Figure~\ref{fig:policy-latency-time} shows the latency behavior over the dynamic workload. Horizontal-only scaling has the highest latency because additional nodes increase coordination overhead in the current model. Vertical-only scaling maintains moderate latency. \textsc{DiagonalScale} produces the lowest average latency by choosing stronger configurations during pressure and cheaper configurations when load drops.

\subsection{Cost Over Time}

\begin{figure}[t]
\centering
\includegraphics[width=\linewidth]{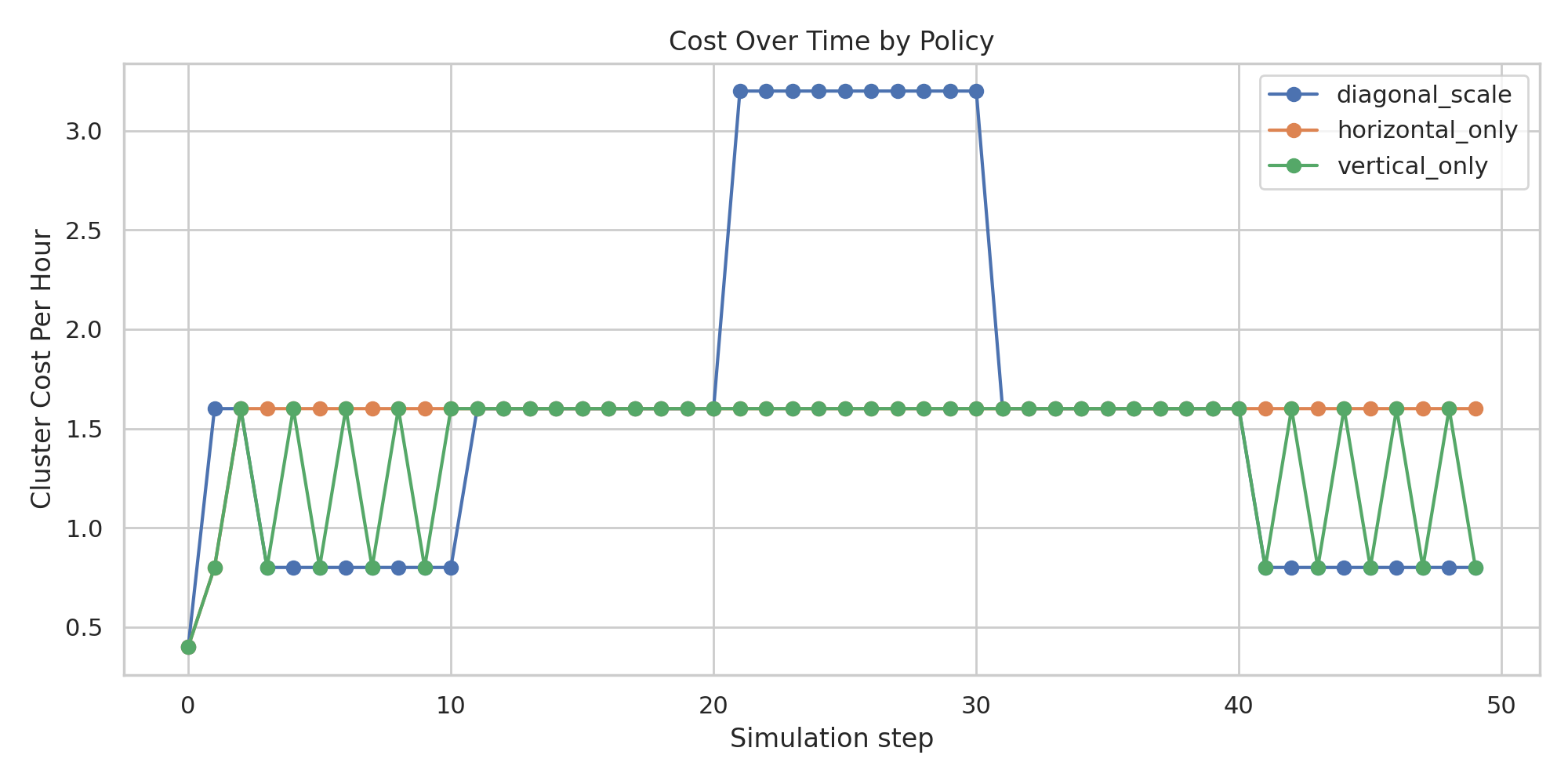}
\caption{Cost over time by policy. \textsc{DiagonalScale} uses higher-cost configurations during the peak workload phase and then scales back down. The result is a slightly higher average cost than the baselines, but substantially better SLA compliance.}
\label{fig:policy-cost-time}
\end{figure}

Figure~\ref{fig:policy-cost-time} shows the cost trajectory. \textsc{DiagonalScale} increases cost during the high-load phase, especially around steps 20--30, but returns to lower-cost configurations afterward. This is a desirable behavior: the policy spends more only when needed to maintain SLA feasibility.

\subsection{Objective Over Time}

\begin{figure}[t]
\centering
\includegraphics[width=\linewidth]{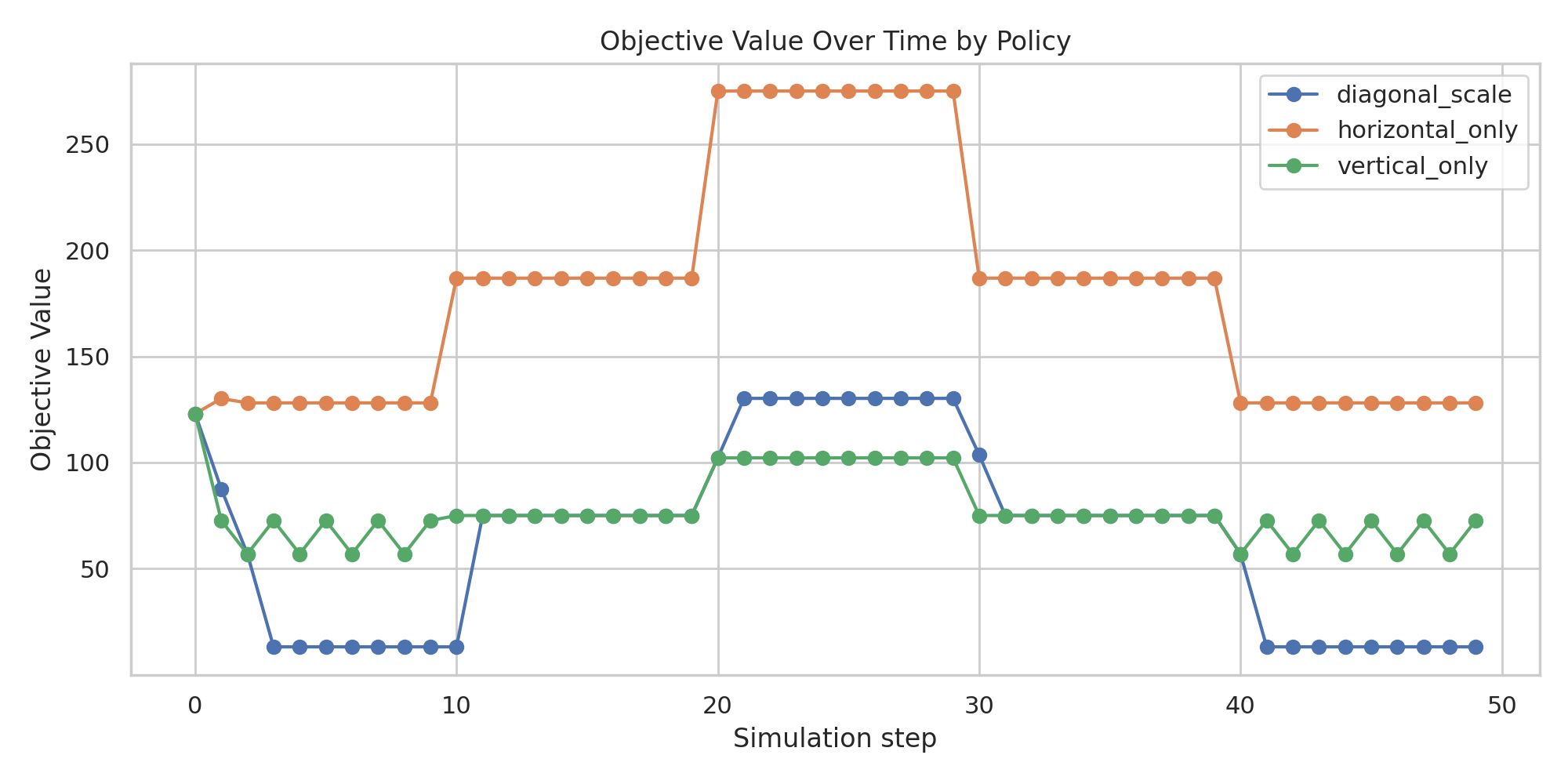}
\caption{Objective value over time by policy. \textsc{DiagonalScale} maintains the lowest average objective value because it balances latency, throughput, coordination cost, cluster cost, and SLA feasibility.}
\label{fig:policy-objective-time}
\end{figure}

Figure~\ref{fig:policy-objective-time} shows that \textsc{DiagonalScale} generally maintains a lower objective than horizontal-only and vertical-only policies. The objective increases during high load but remains below the horizontal-only baseline. This supports the claim that considering both dimensions leads to better trade-offs than axis-aligned scaling.

\subsection{Interpretation}

The most important result is the SLA violation reduction. Before adding SLA-aware feasibility filtering, \textsc{DiagonalScale} could choose cheap but under-provisioned configurations. After adding the feasibility filter, it becomes a practical autoscaling policy rather than merely an unconstrained optimizer. This distinction is central: the objective function identifies desirable regions, but SLA constraints determine which regions are valid.

The latest results suggest the following conclusion: diagonal scaling is useful not because diagonal movement is always selected, but because the policy has access to the full local neighborhood. It can choose a horizontal, vertical, or diagonal step depending on the workload and current configuration. This gives it more control than either baseline.

\section{Limitations}
\label{sec:limitations}

The current evaluation is intentionally limited. First, the simulator is analytical rather than empirical. It does not yet run YCSB against a real CockroachDB, Redis Cluster, Cassandra, or YugabyteDB deployment. Second, the latency function is currently dominated by configuration-dependent terms. Although policies change latency by changing configurations, the model does not yet include a utilization-sensitive queueing term such as $1/(1-u)$. Third, the policy uses one-step local search, so it may require multiple timesteps to reach a feasible configuration during sudden spikes. Fourth, the cost model uses simplified synthetic prices rather than a specific cloud provider's price sheet.

These limitations are acceptable for Phase-1 validation because the goal is to test the structure of the model and the behavior of the policy. They also define a clear path for the next stage of research.

\section{Future Work}
\label{sec:future-work}

The next step is to make the simulator more realistic by adding workload-sensitive latency. A natural model is:
\[
u = \frac{T_{req}}{T(H,V)},
\]
\[
L_{final}=L(H,V) \cdot \frac{1}{1-u},
\]
for $0 \le u < 1$. This would cause latency to spike as utilization approaches capacity and would better capture queueing behavior.

A second extension is empirical calibration. Selected points in the Scaling Plane can be benchmarked using YCSB~\cite{cooper2010ycsb} against systems such as CockroachDB, Redis Cluster, Cassandra, or YugabyteDB. The measured values can then replace or calibrate the analytical surfaces.

A third extension is multi-step lookahead. The current policy evaluates only immediate neighbors. A lookahead controller could search two or three steps ahead, reducing transient SLA violations during sudden spikes. Another extension is to learn the surface online using regression or Bayesian updating while retaining the interpretability of the Scaling Plane model.

Finally, future work should evaluate diagonal scaling in serverless and disaggregated architectures, where compute, memory, storage, and network resources may be scaled independently. Such systems may require a higher-dimensional extension of the Scaling Plane.

\section{Conclusion}
\label{sec:conclusion}

This paper introduced the Scaling Plane, a two-dimensional abstraction for reasoning about distributed database elasticity, and \textsc{DiagonalScale}, an SLA-aware local-search autoscaling policy over that plane. The latest Phase-1 analytical simulation evaluates cost, latency, objective, and policy trajectories over a dynamic workload. The results show that \textsc{DiagonalScale} reduces SLA violations to 3 out of 50 steps, compared with 32 for horizontal-only scaling and 21 for vertical-only scaling. It also achieves the best average latency and objective value, while paying a modest cost premium during peak load.

The findings support the central claim of this work: scaling should not be treated as a binary choice between adding nodes and increasing per-node resources. A policy that can reason over both dimensions can make better trade-offs than one constrained to a single axis. The current simulator establishes a foundation for future empirical calibration and real-system validation.

\bibliographystyle{IEEEtran}
\bibliography{refs}

@inproceedings{corbett2013spanner,
  title        = {Spanner: {Google}'s Globally-Distributed Database},
  author       = {Corbett, James C. and Dean, Jeffrey and Epstein, Michael and Fikes, Andrew and Frost, Christopher and Furman, J. J. and Ghemawat, Sanjay and Gubarev, Andrey and Heiser, Christopher and Hochschild, Peter and Hsieh, Wilson and Kanthak, Sebastian and Kogan, Eugene and Li, Hongyi and Lloyd, Alexander and Melnik, Sergey and Mwaura, David and Naghshineh, Shiva and Quinlan, Sean and Rao, Rajesh and Rolig, Lindsay and Saito, Yasushi and Szymaniak, Michal and Taylor, Christopher and Wang, Ruth and Woodford, Dale},
  booktitle    = {Proceedings of the 10th USENIX Symposium on Operating Systems Design and Implementation (OSDI)},
  year         = {2012},
  pages        = {251--264},
  publisher    = {USENIX Association}
}

@inproceedings{decandia2007dynamo,
  title        = {Dynamo: {Amazon}'s Highly Available Key-value Store},
  author       = {DeCandia, Giuseppe and Hastorun, Deniz and Jampani, Madan and Kakulapati, Gunavardhan and Lakshman, Avinash and Pilchin, Alex and Sivasubramanian, Swaminathan and Vosshall, Peter and Vogels, Werner},
  booktitle    = {Proceedings of the 21st ACM Symposium on Operating Systems Principles (SOSP)},
  year         = {2007},
  pages        = {205--220},
  publisher    = {ACM}
}

@article{chang2008bigtable,
  title        = {Bigtable: A Distributed Storage System for Structured Data},
  author       = {Chang, Fay and Dean, Jeffrey and Ghemawat, Sanjay and Hsieh, Wilson C. and Wallach, Deborah A. and Burrows, Mike and Chandra, Tushar and Fikes, Andrew and Gruber, Robert E.},
  journal      = {ACM Transactions on Computer Systems},
  volume       = {26},
  number       = {2},
  pages        = {4:1--4:26},
  year         = {2008},
  publisher    = {ACM}
}

@article{lakshman2010cassandra,
  title        = {Cassandra: A Decentralized Structured Storage System},
  author       = {Lakshman, Avinash and Malik, Prashant},
  journal      = {ACM SIGOPS Operating Systems Review},
  volume       = {44},
  number       = {2},
  pages        = {35--40},
  year         = {2010},
  publisher    = {ACM}
}

@inproceedings{cooper2010ycsb,
  title        = {Benchmarking Cloud Serving Systems with {YCSB}},
  author       = {Cooper, Brian F. and Silberstein, Adam and Tam, Erwin and Ramakrishnan, Raghu and Sears, Russell},
  booktitle    = {Proceedings of the 1st ACM Symposium on Cloud Computing (SoCC)},
  year         = {2010},
  pages        = {143--154},
  publisher    = {ACM}
}

@misc{cockroachdb_docs,
  title        = {CockroachDB: Architecture and Design Overview},
  author       = {{Cockroach Labs}},
  howpublished = {\url{https://www.cockroachlabs.com/docs/}},
  note         = {Accessed 2025-11-22}
}

@misc{yugabyte_docs,
  title        = {YugabyteDB Architecture},
  author       = {{Yugabyte, Inc.}},
  howpublished = {\url{https://docs.yugabyte.com/}},
  note         = {Accessed 2025-11-22}
}

@misc{aurora_serverlessv2,
  title        = {Amazon Aurora Serverless v2: Instant, Auto-Scaling Configuration for Aurora},
  author       = {{Amazon Web Services}},
  howpublished = {\url{https://aws.amazon.com/rds/aurora/serverless/}},
  note         = {Accessed 2025-11-22}
}

@misc{k8s_hpa,
  title        = {Kubernetes Horizontal Pod Autoscaler},
  author       = {{Kubernetes Authors}},
  howpublished = {\url{https://kubernetes.io/docs/tasks/run-application/horizontal-pod-autoscale/}},
  note         = {Accessed 2025-11-22}
}

@misc{k8s_vpa,
  title        = {Kubernetes Vertical Pod Autoscaler},
  author       = {{Kubernetes Autoscaling SIG}},
  howpublished = {\url{https://github.com/kubernetes/autoscaler/tree/master/vertical-pod-autoscaler}},
  note         = {Accessed 2025-11-22}
}

@misc{k8s_cluster_autoscaler,
  title        = {Kubernetes Cluster Autoscaler},
  author       = {{Kubernetes Autoscaling SIG}},
  howpublished = {\url{https://github.com/kubernetes/autoscaler/tree/master/cluster-autoscaler}},
  note         = {Accessed 2025-11-22}
}

@inproceedings{hellerstein2015serverless,
  title        = {The Rise of {Serverless} Computing},
  author       = {Hellerstein, Joseph M. and Faleiro, Jose and Gonzalez, Joseph E. and Schleier-Smith, Johann and Sreekanti, Vikram and Tumanov, Alexey and Wu, Chenggang},
  booktitle    = {Communications of the ACM},
  year         = {2019},
  volume       = {62},
  number       = {12},
  pages        = {44--54}
}

@inproceedings{verma2015large,
  title        = {Large-scale Cluster Management at {Google} with {Borg}},
  author       = {Verma, Abhishek and Pedrosa, Luis and Korupolu, Madhukar G. and Oppenheimer, David and Tune, Eric and Wilkes, John},
  booktitle    = {Proceedings of the 10th European Conference on Computer Systems (EuroSys)},
  year         = {2015},
  pages        = {1--17},
  publisher    = {ACM}
}

@article{armbrust2010view,
  title        = {A View of Cloud Computing},
  author       = {Armbrust, Michael and Fox, Armando and Griffith, Rean and Joseph, Anthony D. and Katz, Randy and Konwinski, Andy and Lee, Gunho and Patterson, David and Rabkin, Ariel and Stoica, Ion and Zaharia, Matei},
  journal      = {Communications of the ACM},
  volume       = {53},
  number       = {4},
  pages        = {50--58},
  year         = {2010},
  publisher    = {ACM}
}

\end{document}